\documentclass{pasj01}
\begin{document}
\SetRunningHead{K. Sadakane et al.}{Abundances in HD 185330 and 3 Cen A}
\Received{}
\Accepted{}

\title{Spectroscopic  abundance analyses of the ${}^{3}$He stars HD 185330 and 3 Cen A} 


\author{Kozo \textsc{Sadakane},\altaffilmark{1}
         and
        Masayoshi \textsc{Nishimura}\altaffilmark{2}}

\altaffiltext{1}{Astronomical Institute, Osaka Kyoiku University, Asahigaoka,  Kashiwara-shi, 
            Osaka   582-8582}
\email{sadakane@cc.osaka-kyoiku.ac.jp}

\altaffiltext{2}{2-6, Nishiyama-Maruo, Yawata-shi, Kyoto  614-8353 }
\maketitle

\begin{abstract}

Abundances of 21 elements in two  ${}^{3}$He stars HD 185330 and 3 Cen A have been 
analysed relative to the well studied  sharp-lined  B3 V star $\iota$  Her. 
Six elements (P, Ti, Mn, Fe, Ni, and Br) are over-abundant in these two peculiar
stars, while six elements (C, O, Mg, Al,  S, and  Cl) are under-abundant.
Absorption lines of the two rarely observed heavy elements Br~{\sc ii}  
 and Kr~{\sc ii}  are detected in both stars and these elements are both over-abundant.
The centroid wavelengths of the Ca~{\sc ii}  infrared triplet lines in these stars
are red-shifted relative to those lines in  $\iota$  Her and the presence'" of heavy isotopes
of Ca (mass number 44 - 46) in these two stars are confirmed.  
In spite of these similarities, there are several remarkable differences in the abundance
pattern between these two stars.   N is under-abundant
in HD 185330, as in many Hg-Mn stars, while it is significantly over-abundant in 3 Cen A.
P and Ga are both over-abundant in 3 Cen A, while only P is  over-abundant and no 
trace of absorption line of Ga~{\sc ii} can be found  in HD 185330.  
Large over-abundances of Kr and Xe are found in both stars, while the abundance ratios
Kr / Xe are significantly different between them (-- 1.4 dex in HD 185330 and +1.2 dex
in 3 Cen A). Some physical explanations are needed to account for these qualitative
differences.

\end{abstract}

\section{Introduction}

${}^{3}$He stars, which show abnormally strong absorption components caused by
the lighter isotope of helium  (${}^{3}$He), belong to a small group of chemically peculiar
B-type (Bp) stars. These stars are also included in the group of He-weak stars.
\citet{sargent1961} discovered the first member of the group 3 Cen A (HD 120709)
by measuring accurate wavelengths of 10 He~{\sc i} lines. They found a good
correlation between the measured wavelength shifts and the isotope shifts for
${}^{3}$He.  The second star $\iota$ Ori B was discovered by \citet{dwo1973}.

\citet{hartoog1979} searched for B stars with enhanced  ${}^{3}$He and  presented lists of  eight  
definite  ${}^{3}$He stars  and three probable stars. 
They analysed profiles of He~{\sc i} lines and found that  ${}^{3}$He  stars have
total He abundances 5 to 20 times lower than normal B-type stars and have 
 ${}^{3}$He / ${}^{4}$He ratios ranging between  0.47 and 2.7. 
They obtained  the fractional  ${}^{3}$He  content (${}^{3}$He / (${}^{3}$He + ${}^{4}$He)) 
in 11 stars and suggested that  the fractional content of  ${}^{3}$He  appears to 
increase with increasing effective temperature among the  ${}^{3}$He  stars.
They further noted that  ${}^{3}$He  stars occupy a narrow strip in the 
log {\it $T_{\rm eff}$}  - log $\it g$ 
plane between  He-strong stars (hot side) and another  group of He-weak stars (cool side) 
which show no evidence of  ${}^{3}$He  in their spectra.

The list of \citet{hartoog1979} includes HD 185330 (HR 7467) as a definite  ${}^{3}$He star. 
HD 185330 is a 6-th magnitude B-type star classified as B5 II-III by \citet{cowley1972}. 
\citet{preston1976} first recognized the star shows  a red-shifted absorption
component to the He~{\sc i} line at 6678 \AA~  which is caused by the lighter isotope of 
helium.  He noted  that no two stars out of the four  ${}^{3}$He stars known in 
1976 (3 Cen A, $\iota$ Ori B, HD 185330, and $\beta$ Ori B) are spectroscopically alike.  He 
further noted that three stars except for HD 185330 are members of visual binary systems
in young stellar associations.

\citet{michaud1979} carried out detailed radiative acceleration calculations for helium in
stellar envelopes of main sequence stars with effective temperatures between 10,000 
and 25,000 K and 
concluded that the observed under-abundances of  ${}^{4}$He can be explained by
diffusion. They also pointed out  the occurrence of ${}^{3}$He  stars in a narrow 
range of effective temperature can also be explained by diffusion.

\citet{stateva1998} carried out analyses of three He~{\sc i} lines (4921, 5876 and 6678 \AA)
 for four stars (HD 58661, HD 172044, HD 185330, and 46 Aql) 
and confirmed the presence of  ${}^{3}$He  in HD 185330. They 
obtained the isotope ratio  ${}^{3}$He / ${}^{4}$He   to be 0.96 in this star.

\citet{bohlen2005}  analysed profiles of  He~{\sc i} lines in  3 Cen A and HD 185330 
and provided  evidences for different vertical stratification profiles of ${}^{3}$He and 
${}^{4}$He  in these stars. He suggested that  ${}^{4}$He  is generally enhanced deep in 
the photospheres but sharply  depleted high in the atmospheres, 
 while  ${}^{3}$He  appears to be enhanced in a thin layer above that of  ${}^{4}$He. 
Recently, \citet{maza2014} carried out  non-LTE computations for the isotopes of ${}^{3}$He   
and  ${}^{4}$He in the Hg-Mn star $\kappa$ Cnc.    They found that non-LTE abundances
of He obtained from He~{\sc i} lines are lower than LTE values and He is found to be 
stratified in the atmosphere of  $\kappa$ Cnc. They showed that although the LTE analysis 
indicates a step-like  profile of the He abundance,  a gradual decrease 
with height is indicated by the non-LTE analysis.

Information of chemical abundances of elements heavier than helium in atmospheres of ${}^{3}$He  
stars should provide various boundary conditions in further studies of these peculiar 
objects. However, detailed abundance analyses of ${}^{3}$He   stars have been 
published for only 
a few stars including  3 Cen A (\cite{jugaku1961}, \cite{castelli1997}, \cite{adelman2000}, 
and \cite{wahlgren2004}) and $\alpha$ Scl (HD 5737) (\cite{lopez2001}, \cite{saffe2014}) . 
Additional analyses are needed to be carried out
for other stars having different physical conditions such as the effective temperature
in order to increase the number of  sample stars before constructing detailed  
theoretical models.  

In this study, we report chemical abundances of 21 elements (including upper 
limits of two elements) in HD 185330, together with 
the  prototypical  ${}^{3}$He  star 3 Cen A. 
Analyses have been carried out relative to the well studied sharp-lined normal B3 V star 
$\iota$ Her (HD 160762) (\cite{kodaira1970}, \cite{peters1970}, \cite{pintado1983}, 
\cite{golriz2017}) using spectral data  observed with the same instrument.
We confirm isotopic shifts in He~{\sc i} lines and detect the presence of heavy isotopes of 
calcium from IR Ca~{\sc ii} triplet lines. 
A detailed comparison of their abundance patterns provides information of previously 
unknown qualitative differences between these two  ${}^{3}$He stars. 


\section{Observational data}

Spectral data of three target stars in the visible spectral range
(from 3690 to 10480 \AA) were obtained with the Echelle Spectro Polarimetric
Device for the Observation of Stars (ESPaDOnS)\footnote{http://www.cfht.hawaii.edu/Instruments/Spectroscopy/Espadons/} 
attached to the 3.6 m telescope of the Canada-France-Hawaii 
Telescope (CFHT) observatory located on the summit of Mauna Kea, Hawaii. 
Calibrated intensity spectral data were extracted 
from the ESPaDOnS archive through Canadian Astronomical Data Centre (CADC). 
The resolving power is  $\it R$ = 65000.  Details of observational data used 
in the present study are summarized in table 1.
After downloading individual spectral data,  we measured apparent wavelengths
of several isolated and unblended absorption lines of C~{\sc ii} and 
N~{\sc ii}, and obtained averaged shift in wavelength with respect to the 
laboratory scale. We converted the wavelength scale 
of individual spectral data into the laboratory scale. 
All of the available spectra are then averaged into a mean spectrum.
Continuum re-fittings were carried out for each spectral order
separately using high order polynomial functions. The signal-to-noise
(SN) ratios measured on final products at around 6000 \AA~ are 800,  700,
and 1300 for HD 185330, 3 Cen A, and $\iota$ Her, respectively.
   

\setcounter {table} {0}\begin{table}
      \caption{Summary of observational data.} \label{first}
      \begin{center}
      \begin{tabular}{cccc}
\hline\hline
 Object  & Obs date & Exposure   & Number of images \\
\         &          &   (sec)     &                 \\
\hline
HD 185330 & 2005:05:19 &  1200     & 1               \\
          & 2010:12:17 &   240     & 4               \\
          & 2010:12:19 &   390     & 4               \\
3 Cen A   & 2005:05:19 &   300     & 2               \\
$\iota$ Her & 2012:06:25 &  60     & 50              \\
\hline
        \end{tabular}

 \end{center}
\end{table}

\section{Analysis}

Atmospheric parameters ({\it $T_{\rm eff}$} and  log  $\it g$) of HD 185330
have been  estimated from  $\it uvby$ and $\beta$ photometric data taken from
\citet{hauck1998}. We use the empirical calibration of the  [c1] and $\beta$ 
method given in \citet{nieva2012} (equations (7) and (9)) and obtain 
{\it $T_{\rm eff}$} = 16350 K and  log  $\it g$  = 3.80. This temperature is
slightly higher than that adopted in \citet{takeda2010} but coincides within the
expected  error. We adopt parameters of $\iota$ Her given in \citet{nieva2012} 
({\it $T_{\rm eff}$} = 17500 K and  log  $\it g$  = 3.80) 
and those of 3 Cen A obtained by \citet{castelli1997} ({\it $T_{\rm eff}$} = 17500 K 
and  log  $\it g$ = 3.80) . 

We use ATLAS9 model atmospheres \citep{kurucz1993}  after  interpolating 
to the selected atmospheric parameters of each object. Analyses of absorption
lines were carried out using the WIDTH9 program, a companion to the ATLAS9.
For three regions containing merged absorption lines (C~{\sc ii} 4267,  Mg~{\sc ii}
4481 and O~{\sc i} 6156 - 6158),  we use a profile fitting program MPFIT which
was developed by Takeda (1995).  The program calculates a theoretical
spectrum in a small spectral region using a specified model 
atmosphere and a list of atomic lines, and iteratively improves the fitting 
with the observation by changing the relevant parameters, such as the abundances
of specified elements and the macro-turbulent velocity.

Data of atomic transition probabilities (log $\it gf$ values)  were  taken 
from the NIST database \citep{kramida2015} whenever available. When no data
are found in the database, we use log $\it gf$ values given in the VALD3 
database \citep{ryabchikova2015} (Mn~{\sc ii}, Fe~{\sc iii}, and Ni~{\sc ii}).
 We can find no data of atomic transition 
probabilities for two ions Ga~{\sc ii}  and Xe~{\sc ii} in these databases 
and use data taken from \citet{hubrig2014} and stellar log $\it gf$ values given in
 \citet{yuce2011}, respectively. 
Effects of isotopic or hyperfine structures for Mn~{\sc ii}, Ga~{\sc ii}, and Hg~{\sc ii} have
not been taken into acount in this study.

The microturbulent velocity  is defined as  the $\it non-thermal$ component
of the local gas velocity in the spectral line formation region of
the stellar atmosphere \citep{cowley1996}. 
However, the validity of this concept
has been the subject of debate for decades. For instance, it has
been suggested that the use of non-LTE in spectral analysis should
eliminate the need for microturbulence. 
\citet{nieva2012}  carried out extensive
non-LTE analyses of high-resolution optical spectra for a sample
of early B-type stars including  $\iota$  Her. Through an iterative
process, they have constrained the stellar parameters including the
microturbulence. They find a microturbulence of 1 km s${}^{-1}$  in $\iota$
 Her. We have adopted their result of the  microturbulence in  $\iota$ Her in the present
study. The same values of microturbulent velocities are assumed in analyses
of the other two stars HD 185330 and  3 Cen A.  

\begin{figure}
     \begin{center}
       \FigureFile(90mm,100mm){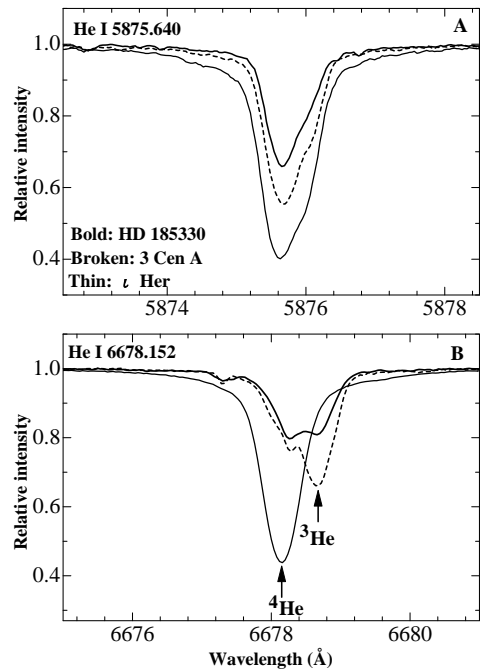}
     \end{center}
     \caption{Comparisons of the He~{\sc i} lines at 5875.64 \AA~ (panel A) and at 6678.15 \AA~
                (panel B) for HD 185330 (bold line), 3 Cen A (broken line) and $\iota$ Her (thin line).
                Absorption centers of isotopes  ${}^{3}$He and  ${}^{4}$He are shown by arrows in 
   panel B. }  
    \end{figure}
\begin{figure}
     \begin{center}
       \FigureFile(90mm,100mm){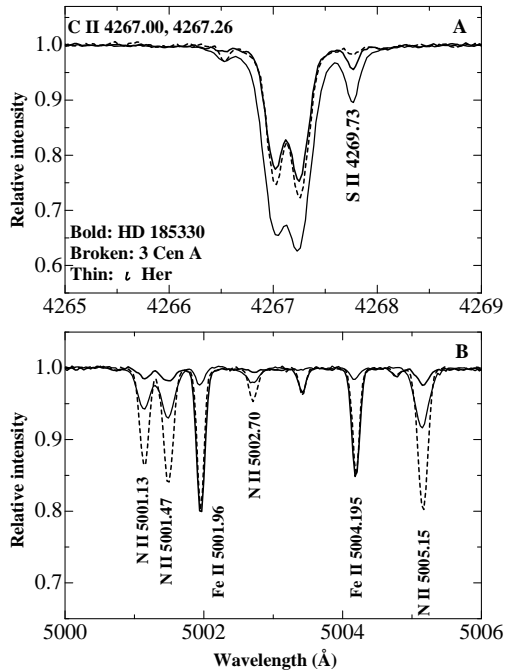}
     \end{center}
     \caption{Comparisons of the C~{\sc ii} doublet lines near 4267\AA~ (panel A) and 
                N~{\sc ii}  lines near 5000 \AA~ (panel B) in three target stars. }  
    \end{figure}
\begin{figure}
     \begin{center}
       \FigureFile(90mm,100mm){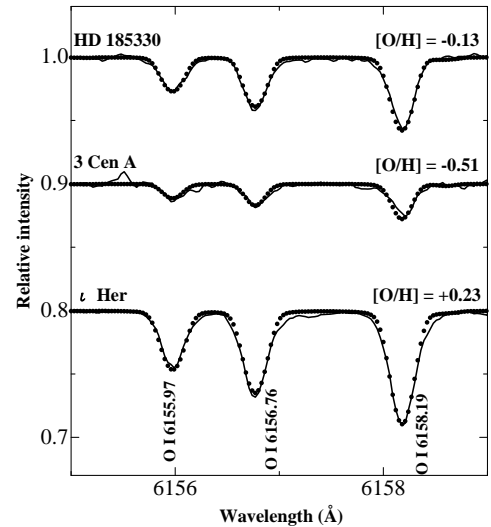}
     \end{center}
     \caption{Analyses of three O~{\sc i} lines near 6157 \AA~ in three target stars.
              Observed profiles (dotted line) and the best fit computed profiles 
              (continuous line) are shown. Best fit LTE abundances of O are indicated
               for each star. }  
    \end{figure}
\begin{figure}
     \begin{center}
       \FigureFile(90mm,100mm){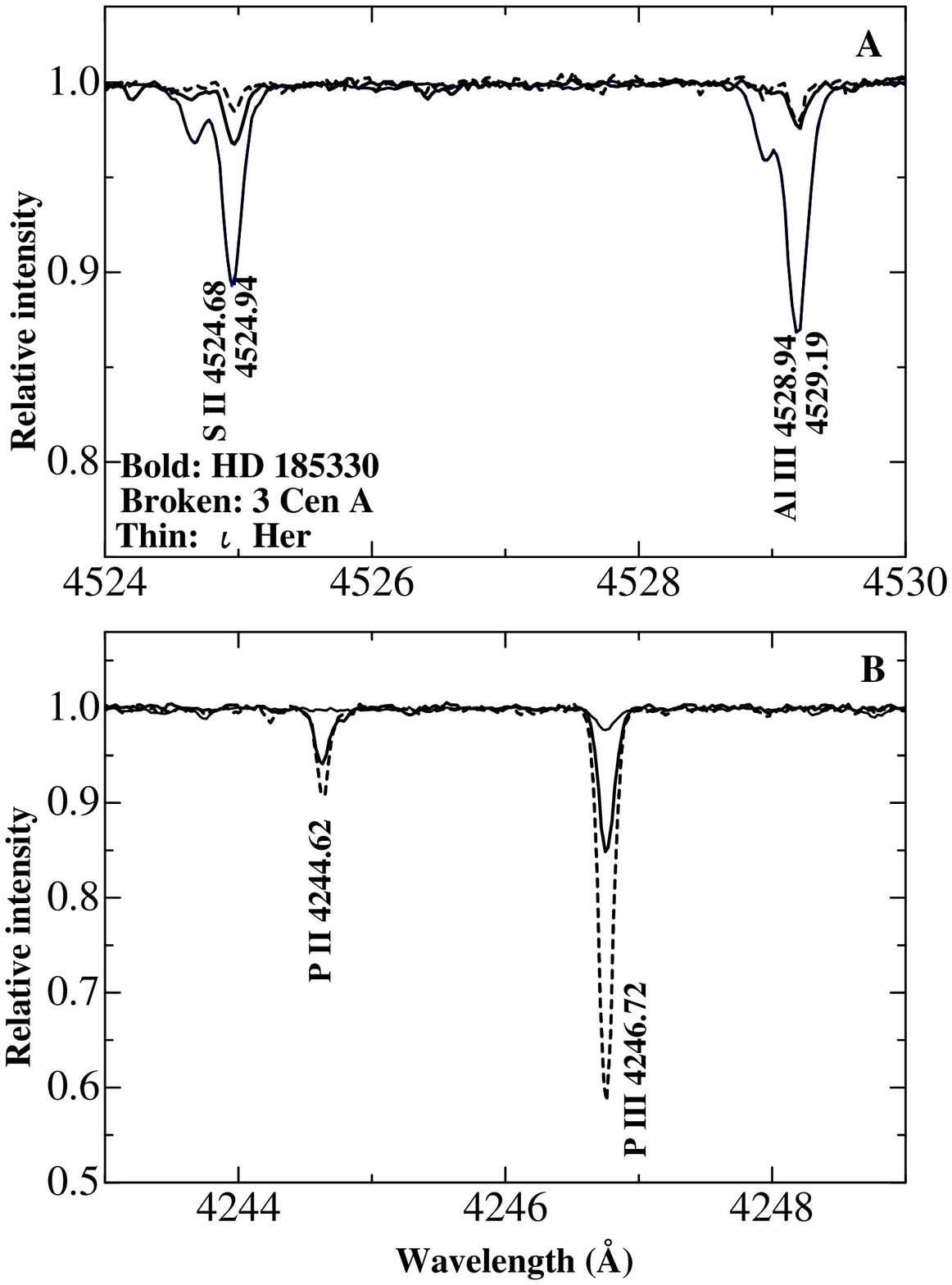}
     \end{center}
     \caption{Comparisons of S~{\sc ii} and Al~{\sc iii} lines (panel A) and P~{\sc ii} and
              P~{\sc iii} lines (panel B) in three target stars.}  
    \end{figure}

\section{Results}

\subsection{Abundances}

We have analyzed absorption lines of 21 elements and obtained new 
abundances for 19 elements and upper limits of two elements (Ga and Hg) in HD 185330.
Abundances of 21 and 15 elements have been obtained in 3 Cen A and  $\iota$ Her, 
respectively.
 We confirm the lighter isotope of He and heavier isotopes of Ca in the two peculiar stars. 
Detailed line-by-line analyses are listed in table 2 (Supplementary data: not shown in this material) . 
Resulting mean abundances of HD 185330 
and 3 Cen A are given in columns 2 and 6 of table 3 and those of the
reference star $\iota$ Her are given in column 10 of the table. 

The main results for HD 185330 are the large over-abundances of 
P [+ 1.9]\footnote{Logarithmic abundance of an element relative to the solar
abundance of the element. [{\it $N_{\rm elem}$}] = log ({\it $N_{\rm elem}$} / 
{\it $N_{\rm H}$})${}_{\rm star}$ -- log ({\it $N_{\rm elem}$} / {\it $N_{\rm H}$})${}_{\rm Sun}$.},
Mn [+ 1.0], Fe [+ 0.4],  Ni [+ 0.5],  Br [+ 2.3], Kr [+ 2.2], and Xe [+ 3.6], 
 and under-abundances of C [-- 0.6],  N [-- 0.5], Mg [-- 0.5],  Al [-- 1.1],   
S [-- 0.9], and Cl [-- 0.5]. Upper limits of two elements (Ga and Hg)  are obtained.
Results of 3 Cen A are over-abundances of N [+ 0.4],  P [+ 2.0],  Mn [+ 1.6],  
Fe [+ 0.4], Ni [+ 0.5], Ga [+ 3.3], Br [+ 2.7], Kr [+ 3.3],  Xe [+ 2.1]  and Hg [+ 3.3] and 
under-abundances of C [-- 0.8],  O [-- 0.6], Mg [-- 0.5],  Al [-- 1.4],  S [-- 1.4], 
and Cl  [-- 0.7].  In $\iota$ Her, we obtained abundances of 15 elements from C to
Ni and found coincidences with the solar abundance \citep{asplund2009} within
$\pm$ 0.3 dex, except for Fe~{\sc ii}. We obtained a large apparent under-abundance 
of Fe from Fe~{\sc ii} lines in this star. 

Because the abundances found for HD 185330 and 3 Cen A are
quite different from the solar abundances \citep{asplund2009}, we computed 
plane parallel ATLAS12 models\footnote{Programs were downloaded from
http://www.astro.utoronto.ca/~lester/programs.html}
 for both stars with the same atmospheric parameters. Abundances
of individual elements obtained above are used in computations. Next, we
tried to obtain $\it new$ abundances with ATLAS12 models using the same
equivalent widths and log $\it gf$ values.
Resulting ATLAS12 abundances for Fe~{\sc ii} and Fe~{\sc iii} show small apparent 
changes ($\sim$ + 0.08 dex and $\sim$  -- 0.02 dex for  Fe~{\sc ii} and Fe~{\sc iii}, respectively) 
from those obtained using ATLAS9 models   
in both stars.  These changes are comparable  to errors resulting
from observational noise or those coming from errors in log $\it gf$ values and in the
adopted atmospheric parameters. Based on these experiments, we decided to adopt 
results obtained with ATLAS9 models in the present study.

We find several large (qualitative) differences in abundances between
these two stars.  N is under-abundant in HD 185330, while it is significantly 
over-abundant in 3 Cen A.
P and Ga are both over-abundant in 3 Cen A, while only P is  over-abundant but no
trace of absorption line of Ga II can be found  in HD 185330.  We notice a large
difference in the abundance ratio Kr / Xe between the two stars. No signature of the 
Hg~{\sc ii} line at 3983 \AA~ is  found in HD 185330, while the Hg~{\sc ii} line is
clearly present and it  is over-abundant in 3 Cen A. 

\noindent
Below we discuss briefly on individual elements.

\noindent
$\it He$: Profiles of two He~{\sc i} lines at 5875.64 \AA~ and at 6678.15 \AA~
 in three target stars HD 185330, 3 Cen A, and $\iota$ Her are compared in figures 1A 
(upper panel) and 1B (lower panel), respectively. The line at  6678 \AA~ clearly 
 shows absorption component  of the lighter isotope   ${}^{3}$He. 
 In 3 Cen A, the ${}^{3}$He component is stronger than that of ${}^{4}$He,
while the ${}^{3}$He component is weaker than that of ${}^{4}$He in HD 185330. 
We have not carried out detailed analyses of the He abundances or the isotope 
ratios  ${}^{3}$He / ${}^{4}$He  and refer to works reported in  \citet{hartoog1979} 
and in \citet{stateva1998}. 

\noindent
$\it C, N$ and $\it O$: Observed profiles of C~{\sc ii} lines at 4267.00 \AA~ and at 4267.26 \AA~
and N~{\sc ii} lines near 5000 \AA~ are compared in figures 2A  and 
2B, respectively. The C~{\sc ii}   lines in HD 185330 and in 3 Cen A are 
both weaker than in $\iota$ Her and we conclude that C is definitely under-abundant
in both stars.  On the other hand,  N~{\sc ii} lines are remarkably weaker in HD 185330
than in 3 Cen A. A difference in the abundance  of N by nearly 1 dex is concluded 
between these two stars.
Line profile  analyses of the three O~{\sc i} lines between 6155 \AA~ and 6159 \AA~
 are displayed in figure 3. When we apply the non-LTE corrections of $\sim$ -- 0.2 dex
for O~{\sc i} given in \citet{takeda2010}, we obtain 
under-abundances of O in HD 185330 [-- 0.3] and in 3 Cen A [-- 0.8],  and
the solar abundance of O in $\iota$ Her. \citet{wahlgren2004} reported an over-abundance
of O [+ 0.15]  from O~{\sc i}  and an under-abundance [-- 0.5] from O~{\sc ii} lines
in 3 Cen A. In these line profile analyses, we assume a rotational velocity ($\it v$ sin $\it i$)
of 5 km s${}^{-1}$ in all three stars. Considering the resolving power of observational data,
we conclude that rotational velocities of  three target stars are equal to or smaller than 
$\sim$ 5 km s${}^{-1}$.
This is in agreement with  the result for $\iota$ Her  ($\it v$ sin $\it i$ = 6 $\pm$ 1 km 
s${}^{-1}$)  given in \citet{nieva2012}.  \citet{sigut2000} reported a very low rotational
velocity ($\it v$ sin $\it i$ $\le$  2 km s${}^{-1}$)  using high resolution
($\it R$ $\sim$ 120000) spectral data.  
\citet{bohlen2005} obtained low rotational velocities ($\it v$ sin $\it i$ = 3 km s${}^{-1}$)  
in both HD 185330 and in 3 Cen A. The rotational velocity  of HD 185330  is lower than that 
 ($\it v$ sin $\it i$ = 15 km s${}^{-1}$) given in \citet{abt2002}.

\noindent
$\it Ne, Mg, Al$ and $\it Si$:  Equivalent widths of 21 lines of Ne~{\sc i} are used 
 to derive LTE abundances of Ne in three target stars. We obtain an apparent over-abundance 
of $\sim$ [+ 0.6] in $\iota$ Her.  Using  non-LTE corrections given 
in \citet{takeda2010}
for the two  Ne~{\sc i}  lines at 6143.06 \AA~ and 6163.59 \AA~ for HD 185330 and $\iota$
Her, we  obtain slight over-abundances of Ne  [+ 0.4]  and [+ 0.3] in these two stars.
Abundances of Mg are obtained from five Mg~{\sc ii} lines including the 4481 \AA~
doublet. Our results show nearly the same under-abundances of Mg  [-- 0.45] in 
both stars.
Profiles of an Al~{\sc iii} line at 4529.19 \AA~ in three target stars are compared 
in figure 4A. We can see the Al~{\sc iii} line is strikingly weak in the 
two peculiar stars. Abundances of Al have been derived from four Al~{\sc iii} lines
and Al is under-abundant by $\sim$ [-- 1.4] in both stars.
Abundances of Si are derived  from  seven Si~{\sc ii} and seven 
Si~{\sc iii} lines. Si shows normal abundance [0.0]  in both HD 185330 and 3 Cen A.

\noindent 
$\it P$: Both HD 185330 and 3 Cen A show very strong lines of P~{\sc ii} and
P~{\sc iii} as illustrated in figures 4B and 7B.  We use 33
lines of P~{\sc ii} and four lines of P~{\sc iii} in the abundance analyses.
We obtain large over-abundances of P  [+ 1.9]  in HD 185330 and
[+ 2.0]  in 3 Cen A, from P~{\sc ii} lines.  A significantly lower abundance of P is obtained 
from P~{\sc iii} lines in HD 185330.  We notice large mean square errors in the resulting
abundances from P~{\sc ii} lines for both stars (around $\pm$ 0.3 dex). 
Furthermore,  abundances of P obtained from P~{\sc ii} lines show apparent 
dependences on equivalent widths in both HD 185330 and 3 Cen A.
We tried to reduce the dependence by changing the value of the microturbulent velocity
from 1 km s${}^{-1}$ to 5 km s${}^{-1}$, and found that the dependence could not be
eliminated. When a large microturbulent velocity is used,  mean square errors in the
abundances of other elements such as Si~{\sc ii} and Fe~{\sc ii} become too large. 

\noindent
$\it S, Cl$ and $\it Ar$: We find 13 S~{\sc ii} lines in HD 185330. Comparisons
of S~{\sc ii} lines in three target stars can be found in figures 2A, 
4A  and 5A. S~{\sc ii} lines in HD 185330 and 3 Cen A are 
weaker than in $\iota$ Her and we obtain under-abundances  [-- 0.9]  and [-- 1.4]
 in these two stars, respectively. We identified two Cl~{\sc ii} lines and
an example is displayed in figure 5A. Cl~{\sc ii} lines are weaker than
in $\iota$ Her and we obtain under-abundances [-- 0.5]  and [-- 0.7]  in 
HD 185330 and 3 Cen A, respectively. Two Ar~{\sc ii} lines at 4426.00 \AA~ and at
4439.19 \AA~ are displayed in figure 5B.  Ar~{\sc ii} lines
in 3 Cen A show  nearly the same strengths as in $\iota$ Her, while Ar~{\sc ii} 
lines in HD 185330 are weaker. Using eight  Ar~{\sc ii} lines, we obtain 
under-abundance of Ar [-- 0.2]  in HD 185330 and normal abundances [0.0] in both
3 Cen A and in $\iota$ Her.  \citet{castelli1997} reported normal abundances 
of S, Cl, and Ar in 3 Cen A. \citet{adelman2000} obtained an under-abundance of
S [-- 1.4]  in this star, which is coincident with our result.

\begin{figure}
     \begin{center}
       \FigureFile(90mm,100mm){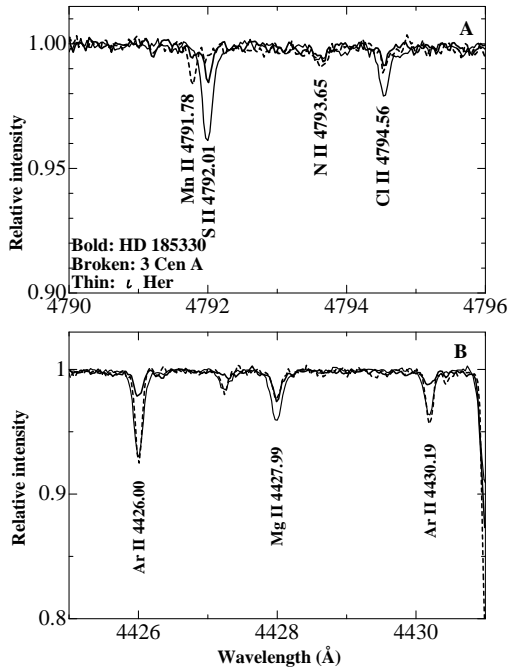}
     \end{center}
     \caption{Comparisons of S~{\sc ii} and Cl~{\sc ii} lines (panel A) and Ar~{\sc ii} and
              Mg~{\sc ii} lines (panel B) in three target stars.}  
    \end{figure}
\begin{figure}
     \begin{center}
       \FigureFile(90mm,100mm){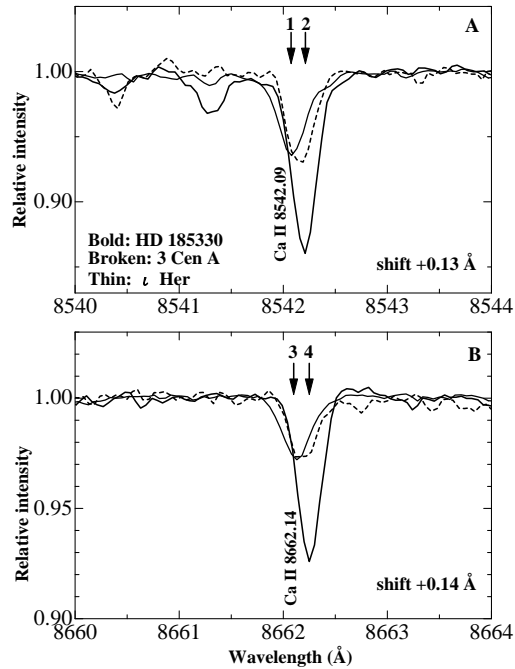}
     \end{center}
     \caption{Comparisons of two Ca~{\sc ii} line at 8542.09 \AA~ (panel A) and at 
              8662.14 \AA~ (panel B) in three target stars. Arrows 1, 3 and 2, 4 show 
              observed line centers in $\iota$ Her and HD185330 for the two Ca~{\sc ii}
              lines, respectively.}  
    \end{figure}
\begin{figure}
     \begin{center}
       \FigureFile(90mm,100mm){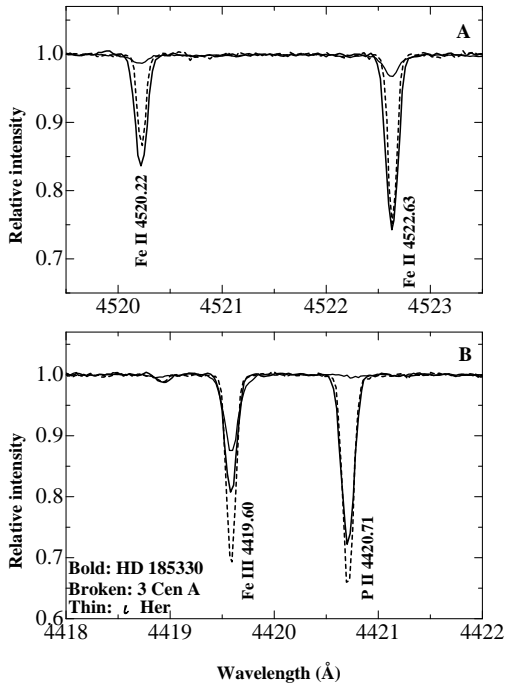}
     \end{center}
     \caption{Comparisons of the two Fe~{\sc ii} lines near 4521 \AA~ (panel A) and an Fe~{\sc iii} and
             a P~{\sc ii} lines near 4705 \AA~ (panel B) in three target stars.}  
    \end{figure}
\begin{figure}
     \begin{center}
       \FigureFile(90mm,100mm){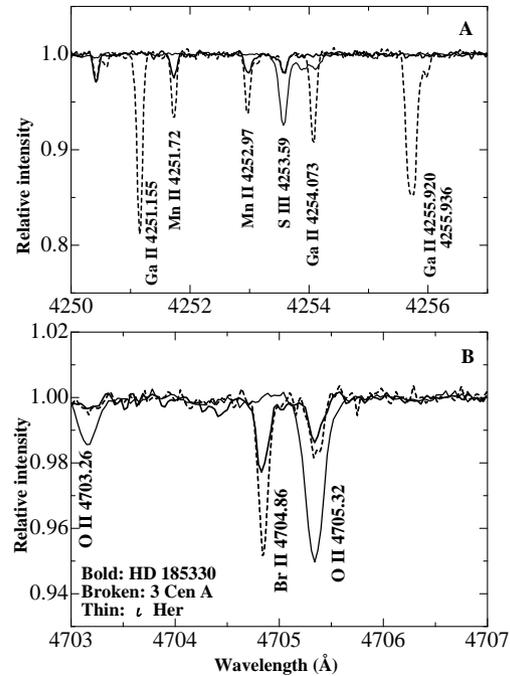}
     \end{center}
     \caption{Comparisons of three Ga~{\sc ii}  lines  near 4254 \AA~ (panel A) and a 
              Br~{\sc ii} line at 4704.86 \AA~ (panel B) in three target stars.}  
    \end{figure}
\setcounter {table} {2}
\begin{table*}
      \caption{Summary of abundances.}\label{third}
\small
      \begin{center}
      \begin{tabular}{cccccccccccccc}
\hline\hline
Ion	&		&	HD 185330	&		&		&	&		& 3 Cen A	&		&		&	$\iota$ Her	&		&    &	Solar	\\
	&	log (N)	&	RMS	&	N	& [N/H]  &	log (N)	& RMS		&	N	& 	[N/H]	& log (N)	&	RMS	&	N	& [N/H]	&	\\
\hline
C~{\sc ii}$^{a}$ 	&	7.82 	&	0.04	&	6	&	-0.61 	&	7.68 	&	0.04 	&	6	&	-0.75 	&	8.38 	&	0.03	&	4	&	-0.05 	&	8.43 	\\
N~{\sc ii}	&	7.32 	&	0.16	&	18	&	-0.51 	&	8.27 	&	0.18 	&	18	&	0.44 	&	7.86 	&	0.07 	&	19	&	0.03 	&	7.83 	\\
O~{\sc i}$^{a}$	&	8.56 	&	0.04	&	3	&	-0.13 	&	8.18 	&	0.04	&	3	&	-0.51 	&	8.92 	&	0.04	&	3	&	0.23 	&	8.69 	\\
O~{\sc ii}	&	8.28 	&	0.14	&	16	&	-0.41 	&	8.11 	&	0.10 	&	16	&	-0.58 	&	8.80 	&	0.07 	&	12	&	0.11 	&	8.69 	\\
Ne~{\sc ii}	&	8.73 	&	0.14	&	21	&	0.80 	&	8.50 	&	0.12 	&	21	&	0.57 	&	8.54 	&	0.20 	&	21	&	0.61 	&	7.93 	\\
Mg~{\sc ii}$^{a}$	&	7.13 	&	0.13	&	7	&	-0.47 	&	7.10 	&	0.15 	&	7	&	-0.50 	&	7.60 	&	0.15 	&	7	&	0.00 	&	7.60 	\\
Al~{\sc ii}	&	5.32 	&	0.28	&	4	&	-1.13 	&	5.08 	&	0.14 	&	4	&	-1.37 	&	6.58 	&	0.11 	&	3	&	0.13 	&	6.45 	\\
Si~{\sc ii}	&	7.39 	&	0.3	&	7	&	-0.12 	&	7.36 	&	0.22 	&	5	&	-0.15 	&	7.50 	&	0.12 	&	5	&	-0.01 	&	7.51 	\\
Si~{\sc iii}	&	7.39 	&	0.11	&	4	&	-0.12 	&	7.65 	&	0.08 	&	4	&	0.14 	&	7.92 	&	0.15 	&	4	&	0.41 	&	7.51 	\\
P~{\sc ii}	&	7.28 	&	0.28	&	33	&	1.87 	&	7.42 	&	0.27 	&	33	&	2.01 	&	5.45 	&	0.25 	&	9	&	0.04 	&	5.41 	\\
P~{\sc iii}	&	6.63 	&	0.22	&	4	&	1.22 	&	7.36 	&	0.18 	&	4	&	1.95 	&	5.40 	&	0.04 	&	3	&	-0.01 	&	5.41 	\\
S~{\sc ii}	&	6.25 	&	0.19	&	13	&	-0.87 	&	5.71 	&	0.19 	&	12	&	-1.41 	&	7.26 	&	0.13 	&	11	&	0.14 	&	7.12 	\\
Cl~{\sc ii}	&	4.98 	&	....	&	2	&	-0.52 	&	4.77 	&	....	&	2	&	-0.73 	&	5.25 	&	....	&	2	&	-0.25 	&	5.50 	\\
Ar~{\sc ii}	&	6.22 	&	0.26	&	8	&	-0.18 	&	6.44 	&	0.14 	&	8	&	0.04 	&	6.69 	&	0.24 	&	8	&	0.29 	&	6.40 	\\
Ti~{\sc ii}	&	5.30 	&	0.14	&	11	&	0.35 	&	5.27 	&	0.17 	&	6	&	0.32 	&	4.79 	&	0.12 	&	6	&	-0.16 	&	4.95 	\\
Cr~{\sc ii}	&	5.43 	&	0.14	&	4	&	-0.21 	&	5.72 	&	0.09 	&	4	&	0.08 	&	5.57 	&	0.04 	&	3	&	-0.07 	&	5.64 	\\
Mn~{\sc ii}	&	6.47 	&	0.17	&	4	&	1.04 	&	7.03 	&	0.04 	&	4	&	1.60 	&	....	&	....	&	....	&	....	&	5.43 	\\
Fe~{\sc ii}	&	7.95 	&	0.16	&	33	&	0.45 	&	7.92 	&	0.14 	&	31	&	0.42 	&	7.13 	&	0.12 	&	19	&	-0.37 	&	7.50 	\\
Fe~{\sc iii}	&	7.75 	&	0.11	&	12	&	0.25 	&	7.86 	&	0.08 	&	9	&	0.36 	&	7.55 	&	0.05 	&	9	&	0.05 	&	7.50 	\\
Ni~{\sc ii}	&	6.70 	&	0.13	&	18	&	0.48 	&	6.68 	&	0.18 	&	18	&	0.46 	&	5.95 	&	0.21 	&	9	&	-0.27 	&	6.22 	\\
Ga~{\sc ii}	&  $\le$ 3.75	&	....	&	2	& $\le$ 0.71	&	6.33 	&	0.21 	&	6	&	3.29 	&	....	&	....	&	....	&	....	&	3.04 	\\
Br~{\sc ii}	&	4.87 	&	0.07	&	3	&	2.33 	&	5.19 	&	0.05 	&	3	&	2.65 	&	....	&	....	&	....	&	....	&	2.54 	\\
Kr~{\sc ii} 	&	5.40 	&	0.02	&	4	&	2.15 	&	6.54 	&	0.08 	&	4	&	3.29 	&	....	&	....	&	....	&	....	&	3.25	\\
Xe~{\sc ii}	&	5.83 	&	0.18	&	9	&	3.59 	&	4.32 	&	.... 	&	2	&	2.08 	&	....	&	....	&	....	&	....	&	2.24 	\\
Hg~{\sc ii}	& $\le$ 3.30	&	....	&	1	&        $\le$	2.13	&	4.47 	&	....	&	1	&	3.30 	& ....	&	....	&	....	&	....	&	1.17 	\\
\hline
        \end{tabular}
 \end{center}
a: Includes results of line profile analyses.
\end{table*}


\noindent
$\it Ca$~{\sc ii}   IR triplet: Profiles of the two Ca~{\sc ii} triplet lines at 8542.09 \AA~
and at 8662.14 \AA~ are displayed in figures 6A and 6B, respectively. 
Both these lines are located near the line centers of H~{\sc i} Paschen
lines (P15 and P13). We adopt the broad wings of the Paschen lines to be the local
continuum and normalized them to unity in these figures.  We have adjusted the wavelength
scale by using the two strong and isolated Ne~{\sc i} lines at 8495.36 \AA~ and at 8654.38 \AA.
We find that the line centers of both Ca~{\sc ii} lines in HD 185330 and 3 Cen A  
are definitely red-shifted relative to those of $\iota$ Her.  Differences in wavelength of
 the observed line centers are + 0.14 \AA~ and + 0.11\AA~ for HD 185330 and 3 Cen A, 
respectively. Using the isotopic shifts of the Ca~{\sc ii} IR triplet lines \citep{nortershauser1998},
we interpret that ${}^{44}$Ca and  ${}^{46}$Ca are dominating in the  atmosphere of HD 185330,
while ${}^{44}$Ca is the mainly contributing isotope in 3 Cen A. 
Inspecting lists of stars in which wavelength shifts of infrared triplet lines of Ca~{\sc ii} 
are observed (\cite{cowley2007} and \cite{cowley2009}),  we notice that 3 Cen A
is presently the hottest star ({\it $T_{\rm eff}$} = 17500 K) and the second hottest star is 
the HZB star Feige 86 ({\it $T_{\rm eff}$} = 16400 K) .

\noindent
$\it Ti, Cr$ and  $\it Mn$: 
Equivalent widths of 11 lines of Ti~{\sc ii} and four lines and Cr~{\sc ii} are 
measured in HD 185330. A slight over-abundance [+ 0.35] and a slight under-abundance
[-- 0.2]  are obtained for these two elements. In 3 Cen A, our measurements 
resulted in a slight over-abundance [+ 0.3]  and a normal abundance of Ti and Cr, 
respectively. We use four Mn~{\sc ii} lines in both stars to obtain definite over-abundances 
of Mn by [+ 1.0] and [+ 1.6]  in  HD 185330 and in 3 Cen A, respectively.
The abundance of Mn obtained in 3 Cen A nearly coincides with published results
(\cite{castelli1997}: [+1.4],  \cite{adelman2000}: [+1.8], and  \cite{wahlgren2004}: [+1.6]).

\noindent
$\it Fe$ and  $\it Ni$: 
Equivalent widths of 33 low excitation (excitation potential $\le$ 3.23 eV) Fe~{\sc ii} 
lines and those of 12  Fe~{\sc iii} lines are used to obtain the Fe abundance in HD 
185330 and over-abundances of [+ 0.45] and [+ 0.25]  are found from Fe~{\sc ii} 
and Fe~{\sc iii}, respectively. For 3 Cen A, we use 31  Fe~{\sc ii} lines and nine 
Fe~{\sc iii} lines and obtain over-abundances [+ 0.42]  from Fe~{\sc ii} and 
[+ 0.36]  from  Fe~{\sc iii}. The abundance of Fe in 3 Cen A obtained in this 
study is in agreements with those reported earlier (\cite{castelli1997}: [+ 0.2] ,  
 \cite{adelman2000}: $\sim$ [+ 0.25], and   \cite{wahlgren2004}: $\sim$ [+ 0.35]).
We obtain somewhat discordant results of the Fe abundance in $\iota$ Her. Figures 7A  
and 7B illustrate profiles of Fe~{\sc ii} and Fe~{\sc iii} lines used in the abundance
analyses in three target stars and we can see that  the  two Fe~{\sc ii} lines near 4520 \AA~
are very weak in $\iota$ Her.  Figure 2B illustrates profiles of two high excitation 
(10.27 eV) lines of Fe~{\sc ii}. These lines are strikingly  weak in  $\iota$ Her. 
Because of the weakness of  Fe~{\sc ii} lines, we  use 19 lines and  obtain an under-abundance 
[-- 0.37]  and a normal abundance [+ 0.05]  from Fe~{\sc iii}  lines in this star. The difference amounts 
to 0.42 dex, well larger than the  rms errors (smaller than 0.15 dex).
The discrepancy in abundances of Fe derived  from Fe~{\sc ii} and Fe~{\sc iii}   
lines in $\iota$ Her
had been pointed out in several papers (e.g. \citet{kodaira1970}, \citet{peters1970}, 
\citet{pintado1983} and \citet{golriz2017}).  They all concluded that a higher  abundance of  Fe is 
obtained from Fe~{\sc iii}  lines than from Fe~{\sc ii}   lines. 
  
We use 18 Ni~{\sc ii}  lines in both HD 185330 and 3 Cen A to derive the abundance 
of Ni and found Ni is over-abundant by $\sim$ [+ 0.45]  in both stars.  On the other hand,
absorption lines of Ni~{\sc ii}  in $\iota$ Her are very weak and we obtain an 
under-abundance of Ni 
[-- 0.3]  using nine lines. Differences in the Ni abundances by $\sim$ 0.7 dex between
the two peculiar stars and  $\iota$ Her strongly suggest enhancements of Ni in these two
stars. We note some comments on the abundance of Ni in these two stars. 
Hg-Mn stars generally show
under-abundances of Ni. \citet{ghaza2016} lists 48 Hg-Mn stars in which abundances 
of Ni have been determined. Only four stars show over-abundances of Ni among them.
\citet{saffe2014}  reported an under-abundance of Ni [-- 0.9]  in a He-weak star 
$\alpha$ Scl. Under-abundances of Ni have been reported in HD 175640 [-- 0.4] 
\citep{castelli2004b} and in HR 6000 [-- 0.4] \citep{castelli2017}. 
In HD 19400, a normal or a slight over-abundance of Ni have been reported in 
\citet{hubrig2014} and in \citet{alonso2003}, respectively. In the two He-weak stars
HD 34797 and HD 35456, \citet{alonso2003} obtained  slight over-abundances of Ni
$\sim$ [+ 0.3].

\noindent
$\it Ga$:  We find  remarkable differences in the observed strengths of Ga~{\sc ii} 
lines between HD 185330 and 3 Cen A. Figure 8A  shows that three 
Ga~{\sc ii} lines near 4254 \AA~  are prominent in 3 Cen A, while these lines are
absent in HD 185330. From equivalent widths of six Ga~{\sc ii} lines, we obtained 
an over-abundance of Ga [+ 3.3]  in 3 Cen A.  This is in agreement with results
obtained by \citet{hidai-1986} and by \citet{castelli1997} based  on 
 ultraviolet IUE spectra of 3 Cen A.  We set upper limits of the two strong lines of
Ga~{\sc ii}  (at 4262.01 \AA~ and 6334.07 \AA)  in HD 185330 to be 0.5 m\AA~ and obtained an
upper limit of [+ 0.7].  The difference in the abundance of Ga between the two 
stars amounts to at least 2.6 dex.

\begin{figure}
     \begin{center}
       \FigureFile(90mm,100mm){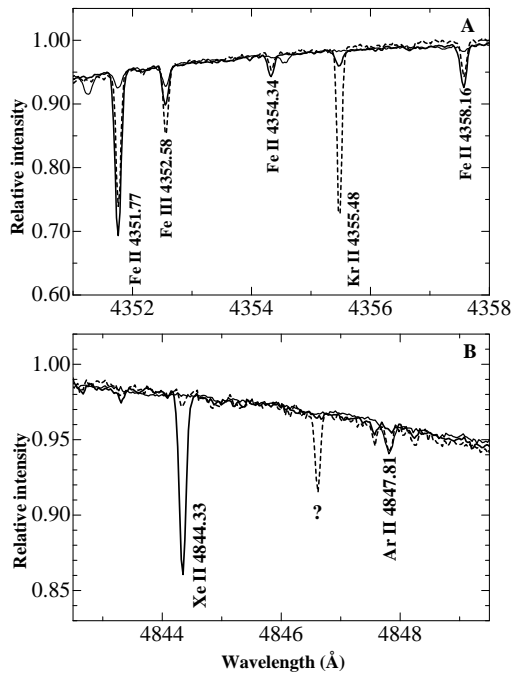}
     \end{center}
     \caption{Comparisons of a Kr~{\sc ii}  line at 4355.48 \AA~ (panel A) and a 
              Xe~{\sc ii} line at 4844.33 \AA~ (panel B) in three target stars.
              A question mark in panel B shows an unidentified line in 3 Cen A.}  
    \end{figure}

\begin{figure}
     \begin{center}
       \FigureFile(90mm,100mm){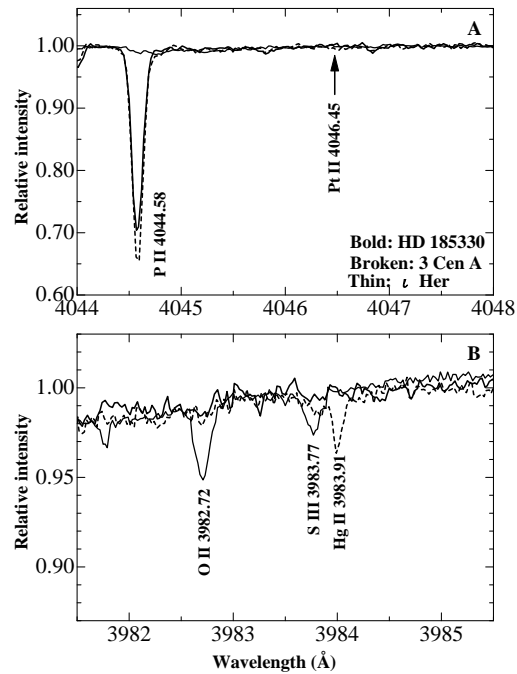}
     \end{center}
     \caption{Comparisons of the Pt~{\sc ii}  line at 4046.45 \AA~ (panel A) and the 
              Hg~{\sc ii} line at 3983.91 \AA~ (panel B) in three target stars. No
              trace of the Pt~{\sc ii} line can be detected.}  
    \end{figure}

\noindent
$\it Br, Kr$ and $\it Xe$:  Absorption lines of the two ions Br~{\sc ii}  and Kr~{\sc ii}  are
rarely observed in stellar spectra. \citet{castelli2004b} analysed two lines of
Br~{\sc ii}  in HD 175640 and obtained an over-abundance of Br  [+ 2.3]  in this
star. \citet{cowley2006} obtained an over-abundance of Br  [+ 2.6]  in 3 Cen A
from three lines of Br~{\sc ii}. Profiles of the Br~{\sc ii} line at 4704.86 \AA~ in
three our target stars are shown in figure 8B. From measurements of 
equivalent widths of three Br~{\sc ii} in HD 185330 and in 3 Cen A, we obtain  
over-abundances of  Br [+ 2.3]  and [+ 2.6]  in these stars, respectively. 
As to Kr~{\sc ii},  \citet{hardorp1968} analysed the
profile of a Kr~{\sc ii} line at 4355.48 \AA~ in 3 Cen A,  and obtained an  
 over-abundance of Kr [+ 3.1].  \citet{castelli1997} obtained  an   over-abundance 
of Kr  [+ 3.3]  from ultraviolet IUE spectra of 3 Cen A. 
Interestingly, we find that  the  Kr~{\sc ii} line at 
4355.48 \AA~   is clearly present but strikingly weaker in HD 185330 than 3 Cen A
(figure 9A). Using equivalent widths of four absorption lines of Kr~{\sc ii},
we obtained over-abundances of [+ 2.2]  and [+ 3.3]  in HD 185330 and 3 Cen A,
respectively.  \citet{castelli1997} reported an over-abundance of Kr [+3.3]  in 3 Cen A
which is in agreement with our result.
The difference in the abundance of Kr between these two stars is 
larger than 1.0 dex.  Profiles of an absorption line of  the next heavy noble gas Xe~{\sc ii} 
at 4844.33 \AA~ is compared in figure 9B. In this case, the  Xe~{\sc ii}  line is
strikingly stronger in HD 185330 than in 3 Cen A. This is just the opposite intensity ratio
to the case of  Kr~{\sc ii} line. From nine clearly visible   Xe~{\sc ii}  lines, we obtained 
an over-abundance of  Xe [+ 3.6]  in HD 185330. In 3 Cen A, we could measure
equivalent widths of only two weak lines and obtained an over-abundance [+ 2.1] 
in this star. 

\noindent
$\it Pt$ and  $\it Hg$:  We have examined the strongest Pt~{\sc ii} line at 4046.45 \AA~
in both HD 185330 and 3 Cen A, but could not find any trace of the line 
(figure 10A). 
The Hg~{\sc ii} line at 3983.93 \AA~ is observed in 3 Cen A, while 
the line is completely absent in both HD 185330 and in $\iota$ Her (figure 10B).  
From the measured equivalent width, we obtain an over-abundance of Hg [+ 3.3] 
in 3 Cen A.  Because of the weakness of the Hg~{\sc ii} line in 3 Cen A, we can say
nothing about the isotopic composition of Hg in this star.
An over-abundance of Hg  [+ 4.5]  is reported in \citet{castelli1997} 
from an analysis of ultraviolet IUE spectra of 3 Cen A. 
In HD 185330, we could obtain an upper limit of [+ 2.1],  assuming an upper limit 
of 0.5 m\AA.

\subsection{Emission lines}

\citet{wahlgren2004}  published extensive lists of emission lines observed in 
3 Cen A. Here, we compare several emission lines in 3 Cen A with those observed 
in HD 185330. 
We compare in figure 11A  emission lines of Mn~{\sc ii} between 6120 \AA~ and 6133 \AA.
These emission lines were discovered in 3 Cen A by \citet{sigut2000} and they are 
clearly present in HD 185330, while most of them are weaker in HD 185330 
than 3 Cen A except for the strongest line at 6122.43 \AA. In $\iota$ Her, a very weak
emission line can be seen at  6122.43 \AA. We notice that no counterpart of the
emission line of Hg~{\sc ii} at 6149.48 \AA~ can be found in HD 185330.
Figure 11B illustrates profiles of the C~{\sc i}  emission line at 9405.73 \AA. 
\citet{alexeeva2016} analysed this emission line in $\iota$ Her and concluded that
their non-LTE calculations can reproduce the observation. The C~{\sc i}  emission 
line is weakly present in both HD 185330 and 3 Cen A.
Profiles of an Fe~{\sc ii}  line at 7513.16 \AA~ is compared in figure 11C. 
In this case,  the line is observed as an  emission line in $\iota$ Her, while the line 
appears as an absorption line in HD 185330. The line shows double peaked emission
components and a sharp central absorption component in 3 Cen A. 
We refrain from further discussions on metallic emission lines observed in these stars 
here and will report further data in a forthcoming paper.

\begin{figure}
     \begin{center}
       \FigureFile(95mm,100mm){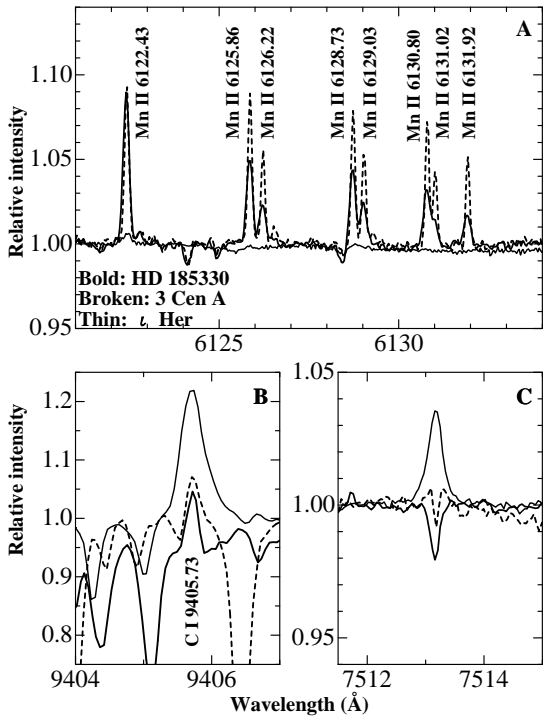}
     \end{center}
     \caption{Weak emission lines in target stars. Panel A shows Mn~{\sc ii} emission lines
              near 6125 \AA. Panels B and C display emission lines of  C~{\sc i} at
              9405.73 \AA~ and of Fe~{\sc ii} at 7513.16 \AA~, respectively. }  
    \end{figure}
\begin{figure}
     \begin{center}
       \FigureFile(90mm,90mm){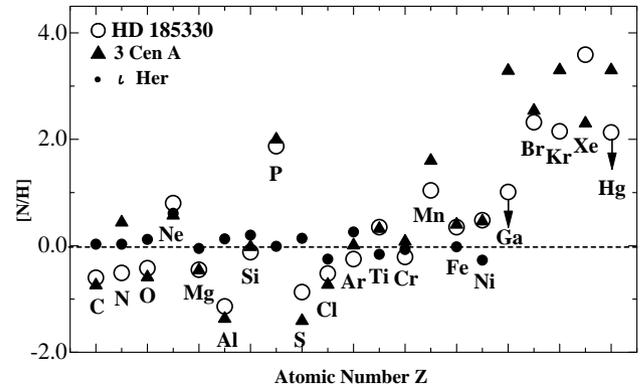}
     \end{center}
     \caption{Abundance patterns of HD 185330 (open circles), 3 Cen A (filled triangles),
              and $\iota$ Her (dots). Abundances relative to the Sun obtained in this
              study are plotted against the atomic number. Downward arrows indicate
              upper limits.}  
    \end{figure}
\begin{figure}
     \begin{center}
       \FigureFile(95mm,100mm){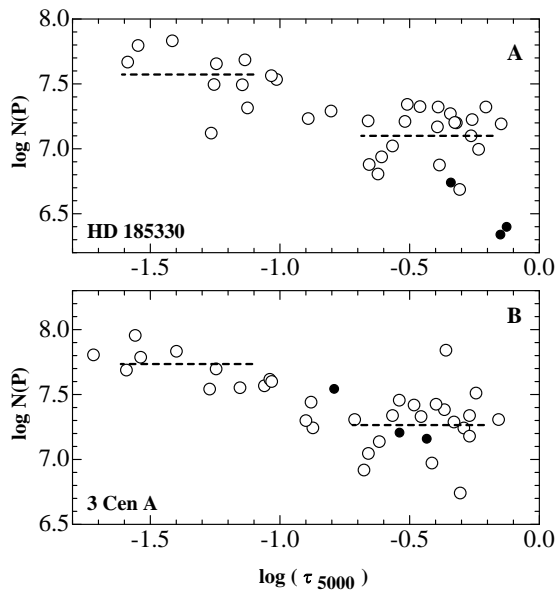}
     \end{center}
     \caption{Possible stratification effects of P in HD185330 (panel A) and 3 Cen A
              (panel B). Open and filled circles show P~{\sc ii} and P~{\sc iii}  lines, 
              respectively. Horizontal dashed lines indicate approximate mean abundances of P 
               in two layers in stellar atmospheres. }
    \end{figure}

\section{Discussion}

We have determined abundances of 21 elements (including two upper limits)
in  ${}^{3}$He stars HD 185330 and 3 Cen A and those of 15 elements  in 
the reference star $\iota$ Her.  We have confirmed that these two stars 
belong to the  ${}^{3}$He subgroup.  We demonstrated that profiles of the two Ca~{\sc ii} triplet 
lines at 8542.09 \AA~ and at 8662.14 \AA~ are  red-shifted relative to those in 
the reference star $\iota$ Her for the first time,  and  have shown that  heavy isotopes  
of Ca (${}^{44}$Ca and  ${}^{46}$Ca)  are present in the  atmospheres of HD 185330 and  3 Cen A. 
Differences in obtained abundances with
respect to the solar abundances are plotted against the atomic number in figure 12.
Hereafter, we focus on apparent differences in the observed abundances between
HD 185330 and 3 Cen A and the possible stratification effect of P suggested
by observations of P~{\sc ii} lines in these two stars.

Under-abundances of light elements (C, O, Mg, Al, S, and Cl) are observed in both 
HD 185330 and 3 Cen A. The case of N should be noted separately. N shows definite
under-abundance [-- 0.5] in HD 185330, while it shows an opposite anomaly
[+ 0.44] in 3 Cen A. The difference of $\sim$ 1.0 dex is far larger than the 
expected error ($\pm$ 0.2 dex) in the analysis. 
\citet{ghaza2016} lists 14 stars in which abundances of N have been
determined in their catalog of Hg-Mn stars. We find only one star (21 Aql)
with a positive value of [+ 0.47]. However, 21 Aql is included in \citet{yuce2014}
as a B8 II-III normal star and not as a Hg-Mn star. If we exclude 21 Aql, values of
[N/H] of 13 Hg-Mn stars are all negative (ranging from -- 0.3 to -- 2.3).
In a remarkable He-weak star HD 65949, which shows extraordinary strong lines
of Re~{\sc ii}, \citet{cowley2010} reported a deficiency of N [-- 2.32], which is
the lowest among the 13 Hg-Mn stars.
 In a non-magnetic Bp star HR 6000, \citet{castelli2017} obtained an under-abundance 
of N [-- 1.02].
Abundances of N in He-weak stars have been determined in several stars. 
\citet{alonso2003}  obtained over-abundances of N in HD 19400  [+0.35] and 
in HD 35456  [+ 0.75] and \citet{lopez2001} obtained a slight over-abundance
of N in $\alpha$ Scl [+ 0.14]. These He-weak stars are cooler than {\it $T_{\rm eff}$} 
= 14000 K  \citep{alonso2003}. Thus, N appears under-abundant in Hg-Mn stars, while
it shows higher abundance among  He-weak stars. If this is the case, HD 185330
might be an exception among He-weak stars.

We find that the abundances of P in both HD 185330 and 3 Cen A obtained  from  
P~{\sc ii} lines have a dependence 
on the optical depth $\tau$, where lines formed in the upper atmosphere show 
larger abundances than lines formed in the deeper layer.  Abundances  of P obtained
from P~{\sc ii}  and P~{\sc iii} lines are plotted against log ($\tau$${}_{5000}$) in
figure 13A (HD 185330) and 13B (3 Cen A). We can see increasing trends of
 abundances of P toward the upper atmospheric layer in both stars. Nearly the 
 same trends of P abundances from P~{\sc ii}  and P~{\sc iii} lines are found
in HR 6000 \citep{castelli2017}.   They listed five spectroscopic observational 
evidences for the vertical abundance stratification in stellar atmospheres.
We find two signs among the five evidences in HD 185330. They are 
1) violation of LTE ionization balance, and 2) disagreement between the 
abundances derived from strong and weak lines of the same ion. 
The large scatter in the
abundance of P could be resulting from the  vertical abundance stratification of 
P in both  HD 185330 and 3 Cen A.  The stratification effects of Mn and Hg
 in  3 Cen A  had been discussed in \citet{wahlgren2004}.

Next, we  compare the pair of P and Ga, because a small group of PGa stars
sometimes appear in papers (e.g. \cite{hubrig2014}) in which both P and Ga are
observed to be enhanced. We find in the present study that both P and 
Ga are definitely over-abundant in 3 Cen A (P [+ 2.0] and Ga [+ 3.3]) while 
in HD 185330 only P is over-abundant [$\sim$ + 1.8]. We obtain only an 
upper limit of Ga in this star. There is a large difference in abundance of Ga
(at least 2.6 dex) between 3 Cen A and HD 185330. We can find similar examples 
in some slightly cooler Bp stars. \citet{castelli2017} obtained an over-abundance 
of P [+ 2.1] while only a slight  over-abundance of Ga [+ 0.36] in HR 6000.
In contrast to HR 6000, \citet{castelli2004b} obtained a slight  over-abundance of P
 [+ 0.31] and a large over-abundance of Ga [+ 3.7] in HD 175640. Interestingly,
in a PGa star HD 19400, both P and Ga show large over-abundances of P [+ 2.3] 
and Ga [+ 3.8] \citep{hubrig2014}. Thus, we find large scatters in the abundance 
ratio of P to Ga among Bp stars and there must be some controlling  mechanisms 
to account for observations. 
   
Abundances of two rarely observed  halogens Cl and Br have been obtained in
 both HD 185330 and 3 Cen A. 
Cl is under-abundant, [-- 0.5] and [-- 0.7] in HD 185330 and 3 Cen A,
respectively. There are only two Hg-Mn stars in which abundances of Cl have 
been obtained \citep{ghaza2016}. They are $\chi$ Lup A [-- 1.40] and HD 46886
[-- 0.36]. \citet{castelli2017} obtained an under-abundance of Cl in HR 6000
[$\le$ -- 1.20]. Our results of Br abundances are over-abundances of [+ 2.3]
and [+ 2.6]  in HD 185330 and 3 Cen A, respectively.  The result of 3 Cen A
coincides with that given in \citet{cowley2006}. An over-abundance of Br [+ 2.29]
has been reported in HD 175640 \citep{castelli2004b}.

Finally, we focus on abundances of the two heavy  elements Kr and Xe. The abundance of Kr
has been obtained only in one star 3 Cen A \citep{hardorp1968}. We use four
absorption lines of Kr~{\sc ii} and obtained over-abundances of Kr in both stars
([+ 2.2] in HD 185330 and [+ 3.3] in 3 Cen A).  We also obtained over-abundances
of Xe in these two stars ([+ 3.6] in HD 185330 and [+ 2.1] in 3 Cen A).  It is 
interesting to find that the abundance ratios  Kr / Xe are reversed in these two 
stars (figure 12).
The abundance of Kr found in HD 185330 is $\sim$ 1.1 dex lower than that  in 3 Cen A, 
while the abundance of Xe is higher  by $\sim$ 1.5 dex.  
\citet{cowley2010} reported abundances of both Kr [+ 2.9] and Xe [+ 4.4] in the
He-weak star HD65949. This star shows nearly the same abundance ratio of  Kr / Xe 
as in HD 185330.  This interesting  contrast in the abundance trends between Kr and Xe
 in these stars needs a physical interpretation.  However, because we have now only a few sample stars in which abundances of Kr have been reported, further observations are needed before constructing any models.  
 
\vskip 3mm

This research has made use of the SIMBAD database, operated
by CDS, Strasbourg, France. We thank Dr. Y. Takeda for his kind help in surveying a
copy of a reference paper and Dr. K. Kato for his co-operation in computing ATLAS12
model atmospheres.

\end {document}